\documentstyle[12pt]{article}
\begin{document}
\title{The Expansion in Width for Domain Walls in Nematic Liquid Crystals 
in External Magnetic Field 
\thanks{Paper supported in part by KBN grant 2P03B 095 13.} 
}
\author{H. Arod\'z   \\
Institute of Physics, Jagellonian University \\
Reymonta 4, 30-059 Cracow, Poland}
\maketitle
\vspace*{1cm}
\begin{abstract}
The improved expansion in width is applied to curved domain walls in 
uniaxial nematic liquid crystals in external magnetic field.
In the present paper we concentrate on the case of equal 
elastic constants.  We obtain approximate form of the director field up to 
second order in magnetic coherence length. 
\end{abstract}

\vspace*{3.5cm}
Preprint TPJU-26/98    
\pagebreak
  
\section{Introduction }
Liquid crystals are probably the best materials for experimental and 
theoretical studies of topological defects. Variety of defects, 
relatively simple experiments in  which one can observe them, and soundness
of theoretical models of dynamics of relevant order parameters make liquid
crystals unique in this respect. Literature on topological defects in 
liquid crystals is enormous, therefore we do not attempt to review it
here. Let us only point the books \cite{1}, \cite{2}, \cite{3} in which 
one can find lucid introductions to the topic as well as collections of 
references. 

Our paper is devoted to dynamics of domain walls in uniaxial nematic liquid
crystals in an external magnetic field. Static, planar domain walls were
discussed for the first time in \cite{5}. We would like to 
approximately calculate director field of a curved domain wall. We use a 
method, called the improved expansion in width, whose general theoretical
formulation has been given in \cite{5, 6}. Appropriately adapted 
expansion in width can also be applied to disclination lines \cite{7}.  

The expansion in  width is based on the idea 
that transverse profiles of the curved domain wall and of a planar one
differ from each other by small corrections which are due to 
curvature of the domain wall. We calculate these corrections perturbatively.
Formally, we expand  the director field in a
parameter which gives the width of the domain wall, that is the magnetic
coherence length $\xi_m$ in the case at hand, but actually terms in the 
expansion involve dimensionless ratios $\xi_m/R_i$, where $R_i$ are (local)
curvature radia of the domain wall. Therefore, our expansion is expected to
provide a good approximation when curvature radia of the domain wall are 
much larger than the magnetic coherence length. 
For planar domain walls the perturbative
solution reduces to just one term which coincides with a well-known exact 
solution.  As we shall see below, the improved expansion in width is not 
quite straightforward -- there are consistency conditions and rather special 
coordinate system is used -- but that should be regarded as a reflection 
of nontriviality of evolution of  curved domain walls. 
Actually, several first terms in the expansion can be calculated without any
difficulty, and the whole approach looks quite promising. 

In the present paper we consider the simplest and rather elegant 
case of equal elastic constants.
In order to take into account differences of values of the elastic 
constants for real liquid crystals one can use, for example, the following
two strategies: perturbative expansion with respect to  
deviations of the elastic constants from their mean 
value, or the expansion in width generalized to the unequal constants case. 
In the former approach, the equal constant approximate solution obtained in
the present paper can be used as the starting point for calculating 
corrections. The case of unequal elastic constants we will discuss in a 
subsequent paper.  

The plan of our paper is as follows. We begin with general description 
of domain walls in uniaxial nematic liquid crystals in Section 2. Next, 
in Section 3, we introduce a special coordinate system comoving with the 
domain wall. Section 4 contains description of the improved expansion in
width. In Section 5 we discuss consecutive terms in the expansion up to 
the second order in $\xi_m$. Several remarks related to our work are 
collected in Section 5.

\section{ Domain walls in nematic liquid crystals }
In this Section we recall basic facts about domain walls in uniaxial nematic
liquid crystals \cite{1}, \cite{2}. We fix our notation and sketch background 
for the calculations presented in next two Sections. 

We shall parametrize the director field $\vec{n}(\vec{x},t)$ by two angles
$\Theta(\vec{x},t)$, $\Phi(\vec{x},t)$:
\begin{equation}
\vec{n} = \left(
\begin{array}{c}
\sin\Theta \cos\Phi \\ \sin\Theta \sin\Phi \\ \cos\Theta 
\end{array}
\right).
\end{equation}
In this way we get rid of the constraint $\vec{n}^2 = 1.$

We assume that the splay, twist and bend elastic constants are equal
($K_{11} = K_{22} = K_{33} = K$). In this case Frank---Oseen---Z\"ocher 
elastic free energy density can be written in the form 
\begin{equation}
{\cal F}_e = \frac{K}{2} (\partial_{\alpha}\Theta \partial_{\alpha}\Theta
+ \sin^2\Theta  \partial_{\alpha}\Phi \partial_{\alpha}\Phi ).
\end{equation}
Our notation is as follows:  $\alpha=1,2,3$; $\;\partial_{\alpha} = 
\partial / \partial x^{\alpha}$; $\;x^{\alpha} $
are Cartesian coordinates in the usual 3-dimensional space $R^3$;
$\;\vec{x} = (x^{\alpha})$. In formula (2) we have abandoned a surface 
term which is irrelevant for our considerations. 

In order to have stable domain walls it is necessary to apply an external 
magnetic field $\vec{H}_0$ \cite{1}, \cite{2}.
We assume that $\vec{H}_0$ is constant in space and time. Without any loss
in generality we may take 
\[ \vec{H}_0 = \left( 
\begin{array}{c}
0 \\ 0 \\ H_0 
\end{array}
\right).
\]
Then the magnetic field contribution to free energy density of the nematic 
is given by the following formula
\begin{equation}
{\cal F}_m = - \frac{1}{2} \chi_a H_0^2 \cos^2\Theta.
\end{equation}
Here $\chi_a$ is the anisotropy of the magnetic susceptibility. It can be 
either positive or negative. For concreteness, we shall assume that 
$\chi_a > 0.$ Our calculations can easily be repeated if $\chi_a < 0.$  
The ground state of the nematic is double degenerate: $\Theta = 0$
and $\Theta = \pi$ give minimal total free energy density
$ {\cal F} = {\cal F}_e + {\cal F}_m. $
It is due to this degeneracy that stable domain walls can exist.

Dynamics of the director field is mathematically described by the equation
\begin{equation}
\gamma_1 \frac{\partial \vec{n}}{\partial t} + \frac{\delta F}{\delta 
\vec{n}} = 0,
\end{equation}
where 
\[  F = \int d^3x {\cal F}. \] 
$\gamma_1$ is the rotational viscosity of the liquid crystal, and 
$\delta/\delta\vec{n}$ denotes the variational derivative with respect to
$\vec{n}$.  Equation (4)
is equivalent to the following equations for the $\Theta$ and $\Phi$ angles
\begin{equation}
\gamma_1 \frac{\partial \Theta}{\partial t} = 
K \Delta \Theta - \frac{K}{2} \sin(2\Theta) \partial_{\alpha}\Phi
\partial_{\alpha}\Phi - \frac{1}{2} \chi_a H_0^2 \sin(2\Theta),
\end{equation}
\begin{equation}
\gamma_1 \sin^2\Theta \frac{\partial \Phi}{\partial t} = 
K \partial_{\alpha}(\sin^2\Theta \partial_{\alpha} \Phi),
\end{equation}
where $\Delta = \partial_{\alpha}\partial_{\alpha}$.

The domain walls arise when the director field is parallel to the 
magnetic field $\vec{H}_0$ in one part of the space and anti-parallel to it 
in another. In between there is a layer -- the domain wall -- across which  
$\vec{n}$ smoothly changes its orientation from parallel to $\vec{H}_0$ 
to the opposite one, that is $\Theta$ varies from 0 to $\pi$ or vice versa. 
The angle $\Phi$ does not play important role. The Ansatz
\begin{equation}
\Phi = \Phi_0 
\end{equation}
with constant $\Phi_0$ trivially solves Eq.(6). Then, Eq.(5) is the only 
equation we have to solve. In the following we 
assume the Ansatz (7), hereby restricting the class of domain walls we 
consider. It is clear from formula (2) that domain walls  with varying 
$\Phi$ have higher elastic free energy than the walls with constant $\Phi$. 

Let us recall the static planar domain wall \cite{1, 2}. We assume that it 
is parallel to the $x^1 = 0$ plane. Then 
\begin{equation}
\Theta = \Theta_0(x^1), \;\; \Phi_0 = \mbox{const},
\end{equation}
where 
\begin{equation}
\Theta_0 \left.\right|_{x^1 \rightarrow -\infty} = 0, \;\;
\Theta_0 \left.\right|_{x^1 \rightarrow +\infty} = \pi.
\end{equation}
One could also consider an "anti-domain wall" obtained by interchanging 0 and 
$\pi$ on the r.h.s. of boundary conditions (9). Equation (5) is now reduced 
to the following equation
\begin{equation}
K \Theta_0^{''} = \frac{1}{2} \chi_a  H_0^2 \sin(2\Theta_0),
\end{equation}
where ' denotes $d/dx^1$. This equation is well-known in soliton theory 
as the sine-Gordon equation, see e.g., \cite{8}. 
It is convenient to introduce the magnetic coherence length $\xi_m$, 
\begin{equation}
\xi_m =\left( K/ \chi_a H^2_0\right)^{1/2}. 
\end{equation} 
The functions 
\begin{equation}
\Theta_0(x^1) = 2 \arctan(\exp \frac{x^1 - x^1_0}{\xi_m})
\end{equation}
with arbitrary constant $x^1_0$ obey Eq.(10) as well as the boundary
conditions (9).  The planar domain walls  are homogeneous in the $x^1=0$ 
plane. Their transverse profile is parametrized by $x^1$. Width of the wall 
is approximately equal to $\xi_m$, in the sense that for $|x^1 - x^1_0| \gg 
\xi_m$ values of $\Theta_0$ differ from 0 or $\pi$ by exponentially small 
terms. 

The planar domain wall solution of Eqs.(5), (6) contains two arbitrary 
constants: $\Phi_0$ and $x^1_0$. The arbitrariness of  $\Phi_0$ is due to the
assumption that the elastic constants are equal. Then the free energy density
${\cal F}$ is invariant with respect to $\Phi \rightarrow \Phi + 
\mbox{const}$. If the elastic constants are not equal this invariance is 
lost, and in the case of planar domain walls $\Phi_0$ can
take only discrete values $n \pi/2, n=0,1,2,3.$ 
The constant $x^1_0$ appears because of invariance of Eqs.(5), (6)
with respect to the translations $x^1 \rightarrow x^1 + \mbox{const}$.

Notice that  $\Theta_0(x^1_0) = \pi/2$. Hence at $x^1 = x^1_0$ the director
$\vec{n}$ is perpendicular to $\vec{H}_0$. In fact, the boundary conditions
(9) imply that for any domain wall there is a surface on which 
$\vec{n}\vec{H}_0 =0$. Such surface is called the core of the domain wall. 
The magnetic free energy density ${\cal F}_m$ has a maximum on the core.

The planar domain wall (12) plays very important role in our approach. In a 
sense, it is taken as the zeroth order approximation to curved domain walls.
The trick consists in using a special coordinate system comoving with the 
curved domain wall. Such a coordinate system encodes shape and motion of 
the domain wall regarded as a surface in the space. Internal dynamics of 
the domain wall, like details of orientation of the director inside the 
domain wall, is then calculated perturbatively in the comoving reference 
frame with the function (12) taken as the leading term.

\section{The comoving coordinates}
The first step in our construction of the perturbative solution consists 
in introducing the coordinates comoving with the domain wall.
Two coordinates ($\sigma^1,\; \sigma^2$) parametrize the domain wall 
regarded as a surface in the $R^3$ space, and one coordinate, let say $\xi$, 
parametrizes the direction perpendicular to the domain wall. For convenience 
of the reader we quote main definitions and formulas below \cite{6}.

We introduce a smooth, closed or infinite surface $S$ in the usual $R^3$ 
space. It is supposed to lie close to the domain wall. Its shape mimics 
the shape of the domain wall. In particular we may assume that
$S$ coincides with the core at certain time $t_0$. 
Points of $S$ are given by  $\vec{X}(\sigma^i,t)$, where $\sigma^i$ $(i=1,2)$
are two intrinsic coordinates on $S$, and $t$ denotes the time. We
allow for motion of $S$ in the space. The vectors $\vec{X}_{,k},\; k=1,2$,
are tangent to $S$ at the point $\vec{X}(\sigma^i,t)$ \footnote{We use the 
notation $f_{,k} \equiv \partial f / \partial\sigma^k$.}. They are 
linearly independent, but not necessarily orthogonal to each other. At each 
point of $S$ we also introduce a unit vector 
$\vec{p}(\sigma^i,t)$  perpendicular to $S$, that is
\[ \vec{p} \vec{X}_{,k} =0,\;\;\; \vec{p}^2 =1. \]
The triad $(\vec{X}_{,k}, \vec{p})$ forms a local basis at the point 
$\vec{X}$ of $S$. Geometrically, $S$ is characterized by the
induced metric tensor on $S$
\[ g_{ik} = \vec{X}_{,i} \vec{X}_{,k},   \]
and by the extrinsic curvature coefficients of $S$
\[ K_{il} = \vec{p}\vec{X}_{,il}, \]
where $i,k,l=1,2$. They appear in Gauss-Weingarten formulas  
\begin{equation}
\vec{X}_{,ij} = K_{ij} \vec{p} + \Gamma^l_{ij} \vec{X}_{,l}, \;\;\;\;
\vec{p}_{,i} = - g^{jl} K_{li} \vec{X}_{,j}. 
\end{equation}
The matrix $(g^{ik})$ is by definition the inverse of the matrix 
$(g_{kl})$, i.e.
$g^{ik}g_{kl}= \delta^i_l$, and $\Gamma^l_{ik}$ are Christoffel symbols
constructed from the metric tensor $g_{ik}$. Two eigenvalues  $k_1,k_2$ of 
the matrix $(K^i_j)$, where  $K^i_j = g^{il}K_{lj}$, are called extrinsic 
curvatures of $S$ at the point $\vec{X}$. The main curvature radia are 
defined as $R_i =1/k_i$. 

The comoving coordinates $(\sigma^1, \sigma^2,\xi)$ are introduced by the 
following formula
\begin{equation}
\vec{x} = \vec{X}(\sigma^i,t) + \xi \vec{p}(\sigma^i,t).
\end{equation}
$\xi$ is the coordinate in the direction perpendicular to the surface $S$. 
In the comoving coordinates this surface has very simple equation: $\xi = 0.$
We will use the compact notation: 
$(\sigma^1, \sigma^2, \xi) = (\sigma^{\alpha})$, where $\alpha$=1, 2, 3 
and $\sigma^3 = \xi$. The coordinates $(\sigma^{\alpha})$ are just a special 
case of curvilinear coordinates in the space $R^3$. In these coordinates the
metric tensor $(G_{\alpha \beta})$ in $R^3$ has the following components:
\[ G_{33} =1, \;\; G_{3k} = G_{k3} =0, \;\; G_{ik} = N^l_i g_{lr}N^r_k, \]
where 
\[ N^l_i = \delta^l_i - \xi K^l_i, \]
$i,k,l,r =1,2$. Simple calculations give 
\[ \sqrt{G} =\sqrt{g} N, \]
where $G=det(G_{\alpha\beta}), 
\;\; g=det(g_{ik})$ and $ N=det(N_{ik}).$ For $N$ we obtain the
following formula 
\[ N = 1 - \xi K^i_i + \frac{1}{2} \xi^2 (K_i^i K^l_l - K^i_lK^l_i). \]
Components $G^{\alpha \beta}$ of the inverse metric tensor in $R^3$ have 
the form  
\[ G^{33} =1,\;\; G^{3k}=G^{k3}=0, \;\; G^{ik}= (N^{-1})^i_r g^{rl} 
(N^{-1})^k_l, \]
where 
\[ (N^{-1})^i_r = \frac{1}{N} 
\left((1-\xi K^l_l) \delta^i_r + \xi K^i_r \right). \]
We see that dependence on the transverse coordinate 
$\xi$ is explicit, while $\sigma^1, \sigma^2$ appear through the tensors
$g_{ik}, \; K^l_r$ which characterize the surface $S$. 

The comoving coordinates $(\sigma^{\alpha})$ have in general certain finite 
region of validity. In particular, the range of $\xi$ is given by the smallest
positive $\xi_0(\sigma^i,t)$ for which $G = 0$. It is clear that such $\xi_0$ 
increases with decreasing extrinsic curvature coefficients $K_i^l$, reaching
infinity for the planar domain wall, for which $K_j^i = 0$. We assume that 
the surface $S$ (hence also the domain wall) is not curved too much. Then,
that region is large enough, so that outside it  
there are only exponentially small tails of the domain wall which give 
negligible contributions to physical characteristics of the domain wall. 

The comoving coordinates are utilised to write Eq.(5) in a form suitable 
for calculating the curvature corrections. Let us start from the Laplacian 
$\Delta\Theta$. In the new coordinates it has the form 
\[ 
\Delta\Theta = \frac{1}{\sqrt{G}} \frac{\partial}{\partial\sigma^{\alpha}}
\left(\sqrt{G} G^{\alpha\beta}\frac{\partial\Theta}{\partial \sigma^{\beta}}
\right).
\] 
The time derivative on the l.h.s. of Eq.(5) is taken under the condition 
that all $x^{\alpha}$ are constant. It is convenient to use time derivative 
taken at constant $\sigma^{\alpha}$. The two derivatives are related by 
the formula
\[
\frac{\partial}{\partial t}|_{x^{\alpha}} = 
\frac{\partial}{\partial t}|_{\sigma^{\alpha}} +
\frac{\partial \sigma^{\beta}}{\partial t}|_{x^{\alpha}}
\frac{\partial}{\partial \sigma^{\beta}},
\]
where
\[ \frac{\partial \xi}{\partial t}|_{x^{\alpha}} 
= - \vec{p}\dot{\vec{X}}, \;\;\;
\frac{\partial \sigma^i}{\partial t}|_{x^{\alpha}} = - (N^{-1})^i_k
g^{kr} \vec{X}_{,r} (\dot{\vec{X}} + \xi \dot{\vec{p}}), 
\]
the dots stand for $\partial /\partial t |_{\sigma^i}$. 
Let us also introduce the dimensionless coordinate 
\[ s = \xi / \xi_m. \] 
Now we can write equation (5) transformed to the comoving coordinates 
$(\sigma^i, s)$ (with the Ansatz (7) taken into account):
\[
\frac{\gamma_1}{K} \xi_m^2 
\left( \frac{\partial \Theta}{\partial t}|_{\sigma^{\alpha}}
- \frac{1}{\xi_m}\vec{p}\dot{\vec{X}} \frac{\partial\Theta}{\partial s}
- (N^{-1})^i_k g^{kr} \vec{X}_{,r} (\dot{\vec{X}} + \xi_m s
\dot{\vec{p}}) \frac{\partial\Theta}{\partial\sigma^i} \right) 
\]
\begin{equation}
= \frac{\partial^2 \Theta}{\partial s^2} - 
\frac{1}{2} \sin(2\Theta) + \frac{1}{N} \frac{\partial N}{\partial s} 
\frac{\partial \Theta}{\partial s} + \xi_m^2 \frac{1}{\sqrt{g} N} 
\frac{\partial}{\partial \sigma^j} \left(G^{jk}\sqrt{g} N 
\frac{\partial \Theta}{\partial \sigma^k} \right), 
\end{equation}
Equation (15) is the starting point for construction of the expansion 
in width.

\section{The improved expansion in width}
We seek domain wall solutions of Eq.(15) in the form of expansion with 
respect to $\xi_m$, that is 
\begin{equation} 
\Theta = \Theta_0 + \xi_m \Theta_1 + \xi_m^2 \Theta_2 + ... \;\;\;   .
\end{equation}
Inserting formula (16) in Eq.(15) and keeping only terms of the lowest
order ($\sim \xi_m^0$) we obtain the following equation 
\begin{equation}
\frac{\partial^2 \Theta_0}{\partial s^2} = \frac{1}{2} \sin(2\Theta_0),
\end{equation}
which essentially coincides with Eq.(10) after the rescaling $x^1 = \xi_m s$. 
Its solutions 
\[
\Theta_{s_0}(s) = 2 \arctan(\exp(s-s_0)),
\]
essentially have the same form as the planar domain walls (12), but now $s$ 
gives the distance from the surface $S$. This surface will be determined 
later. In the remaining part of the paper we shall consider curvature 
corrections to the simplest solution
\begin{equation}
\Theta_0(s) = 2 \arctan(\exp s).
\end{equation}
Because already $\Theta_0$ interpolates between the ground state solutions 
$0, \pi$, the corrections $\Theta_k, \; k \geq 1,$ should vanish in the 
limits $s \rightarrow \pm \infty$.  

Equations for the corrections $\Theta_k, \; k\geq1,$  are obtained  by 
expanding both sides of Eq.(15) and equating terms proportional
to $\xi_m^k$. These equations can be written in the form 
\begin{equation}
\hat{L} \Theta_k = f_k,
\end{equation}
with the operator $\hat{L}$ 
\begin{equation}
\hat{L} = \frac{\partial^2}{\partial s^2} - \cos(2\Theta_0)  =
\frac{\partial^2}{\partial s^2} + \frac{2}{\cosh^2 s} -1.
\end{equation}
The last equality in (20) can be obtained, e.g., from Eq.(17): 
inserting $\Theta_0$ given by formula (18) on the l.h.s. of
Eq.(17) we find that $ \sin(2 \Theta_0) = - 2 \sinh s/ \cosh^2 s$, and
$\cos(2\Theta_0) = 1 - 2/\cosh^2s. $
The expressions $f_k$ on the r.h.s. of Eqs.(19) depend on the lower order 
contributions $\Theta_l, \; l<k$. Straightforward calculations give 
\begin{equation}
f_1 = \partial_s\Theta_0 ( K^r_r - \frac{\gamma_1}{K} \vec{p}\dot{\vec{X}}),
\end{equation}
\begin{equation}
f_2 = - \sin(2\Theta_0) \Theta_1^2 + s \partial_s\Theta_0 K^i_j K^j_i
+ \partial_s\Theta_1 ( K^r_r - \frac{\gamma_1}{K} \vec{p}\dot{\vec{X}}),
\end{equation}
\begin{eqnarray}
& f_3 = \frac{\gamma_1}{K} (\partial_t\Theta_1 - g^{kr}\vec{X}_{,r}
\dot{\vec{X}} \partial_k\Theta_1 ) - 2 \sin(2\Theta_0) \Theta_1 \Theta_2
- \frac{2}{3} \cos(2\Theta_0) \Theta_1^3  & \nonumber \\
&+ s \partial_s\Theta_1 K^i_jK^j_i - \frac{1}{2} s^2 \partial_s\Theta_0 K^r_r 
\left( (K^i_i)^2 - 3 K^i_jK^j_i\right) & \nonumber \\
& - \frac{1}{\sqrt{g}}\partial_j(\sqrt{g}g^{jk}\partial_k\Theta_1) +
\partial_s\Theta_2 ( K^r_r - \frac{\gamma_1}{K} \vec{p}\dot{\vec{X}}),
\end{eqnarray}
and
\begin{eqnarray}
& f_4 = \frac{\gamma_1}{K} ( \partial_t\Theta_2 
- s g^{ik} \dot{\vec{p}}\vec{X}_{,k}\partial_i\Theta_1 )
- \frac{\gamma_1}{K} g^{jk} \vec{X}_{,k}\dot{\vec{X}} (\partial_j\Theta_2 +
s K^i_j\partial_i\Theta_1 ) & \nonumber \\
& - \sin(2\Theta_0) (\Theta_2^2 +2 \Theta_1 \Theta_3 - 
\frac{1}{3} \Theta_1^4) - 2 \cos(2\Theta_0) \Theta_1^2 \Theta_2 & 
\nonumber \\ 
& + s \partial_s \Theta_2 K^i_jK^j_i  
+ s^3 \partial_s\Theta_0 \left( (K^r_r)^4 + \frac{1}{2} (K^r_sK^s_r)^2 
- 2 (K^r_r)^2 K^i_jK^j_i \right) & \nonumber \\
& - \frac{s^2}{2} \partial_s\Theta_1 K^r_r \left(
(K^i_i)^2 - 3 K^i_jK^j_i \right) 
- \frac{1}{\sqrt{g}}\partial_j(\sqrt{g}g^{jk}\partial_k\Theta_2) 
& \nonumber \\
& -\frac{2s}{\sqrt{g}} \partial_j(\sqrt{g}K^{jk}\partial_k\Theta_1)
+ s g^{jk}(\partial_jK^r_r) \partial_k\Theta_1 +  \partial_s\Theta_3
( K^r_r - \frac{\gamma_1}{K} \vec{p}\dot{\vec{X}}),
\end{eqnarray}
where $\partial_t = \partial /\partial t,\;\; \partial_i = \partial / 
\partial \sigma^i$. We have taken into account the fact that $\Theta_0$ does 
not depend on $\sigma^i$.

Notice that all Eqs.(19) for $\Theta_k$ are linear. The only nonlinear 
equation in our perturbative scheme is the zeroth order equation (17). 

It is very important to observe that operator $\hat{L}$ has a zero-mode, 
that is a function $\psi_0(s)$ which quickly vanishes in the limits 
$s \rightarrow \pm \infty$, and which obeys the equation
\[
\hat{L} \psi_0 =0.
\]
Inserting $\Theta_{s_0}(s)$ in Eq.(17),  differentiating that equation with
respect to $s_0$ and putting $s_0=0$ we obtain as an identity that
$\hat{L}\psi_0 = 0$ where
\begin{equation}
\psi_0(s) = \frac{1}{\cosh s}.
\end{equation}
The presence of this zero-mode is related to the invariance of Eq.(17) 
with respect to translations in $s$, therefore it is often called the 
translational zero-mode. Let us multiply both sides of Eqs.(19) by 
$\psi_0(s)$ and integrate over $s$. Integration by parts gives 
\[
\int^{\infty}_{-\infty} ds \psi_0 \hat{L} \Theta_k =
\int^{\infty}_{-\infty} ds \Theta_k  \hat{L} \psi_0  = 0. 
\]
Hence, we obtain the consistency (or integrability) conditions 
\begin{equation}
\int^{\infty}_{-\infty} ds \psi_0(s) f_k(s) =0,
\end{equation}
where $f_k$ are given by formulas of the type (21) - (24). We shall see in 
the next Section that these conditions play very important role in
determining the curved domain wall solutions. 

Using standard methods \cite{9} one can obtain the following formulas for 
vanishing in the limits $s \rightarrow \pm \infty$ solutions $\Theta_k$ 
of Eqs.(19):
\begin{equation}
\Theta_k = G[f_k] + C_k(\sigma^i, t) \psi_0(s), 
\end{equation}
where
\begin{equation}
G[f_k] = -  \psi_0(s) \int^s_0 dx \psi_1(x) f_k(x) +  \psi_1(s) 
\int^s_{-\infty} dx \psi_0(x) f_k(x). 
\end{equation}
Here $\psi_0(s)$ is the zero-mode (25) and 
\begin{equation}
\psi_1(s) = \frac{1}{2} (\sinh s + \frac{s}{\cosh s}) 
\end{equation}
is the other solution of the homogeneous equation 
\[ 
\hat{L} \psi = 0.
\]
The second term on the r.h.s. of formula (27) obeys the homogeneous 
equation $\hat{L}\Theta_k =0$. It vanishes when $s \rightarrow \pm \infty$. 

The solutions (27) contain as yet arbitrary functions $C_k(\sigma^i, t)$. 
Also $\vec{X}(\sigma^i, t)$ giving the comoving surface $S$ has not been
specified.  It turns out that conditions (26) are so restrictive that they 
essentially fix those functions. The extrinsic curvature coefficients 
$K^i_l$ and the metric $g_{ik}$ will follow from $\vec{X}(\sigma^i,t)$.

One can worry that $G[f_k], \; k\geq 1$, given by formula (28) 
do not vanish when $s\rightarrow \pm \infty$ because the second term on 
the r.h.s. of formula (28) is proportional to $\psi_1$,  which 
exponentially increases in the limits $s \rightarrow \pm \infty$. 
However, the integrals 
\[ \int^s_{-\infty} dx \psi_0 f_k \]
vanish in that limit due to the consistency conditions (26). Moreover, 
qualitative analysis of Eq.(15) shows  that $f_k \sim 
(\mbox{polynomial in}\; s) \times \exp(-|s|)$ for large $|s|$, hence those
integrals behave like $(\mbox{polynomial in}\;s) \times \exp(-2|s|)$ for 
large $|s|$ ensuring that all $G[f_k]$ exponentially  vanish when 
$|s| \rightarrow \infty$.

\section{The approximate domain wall solutions}
In this Section we discuss the approximate solutions obtained with the help
of the perturbative scheme we have just described. We present formulas for 
$\Theta_1$ and $\Theta_2$, an equation for $\vec{X}(\sigma^i, t)$ which
determines motion of the surface $S$, as well as equations for the 
functions $C_1,\;C_2$. 

The zeroth order solution is already known, see formula (18). This allows us
to discuss the consistency condition with  $k=1$. Substituting $f_1$ from 
formula (21) and noticing that 
\[ \partial_s \Theta_0 = \frac{1}{\cosh s} = \psi_0(s) \]
we find that that condition is equivalent to
\begin{equation}
\frac{\gamma_1}{K} \vec{p} \dot{\vec{X}} = K^r_r.
\end{equation}
This condition is in fact the equation for $\vec{X}$. It is of the same 
type as Allen-Cahn equation \cite{10}, but in our approach it governs 
motion of the auxiliary surface $S$. 

Let us now turn to the perturbative corrections. After taking into account 
Allen-Cahn equation (30) we have $f_1 =0$. Therefore, the total
first order contribution has the form
\begin{equation}
\Theta_1 = \frac{C_1(\sigma^i,t)}{\cosh s}.
\end{equation}

The second order contribution $\Theta_2$ is calculated from formula (28) 
with $f_2$ given by formula (22).
Using the results (30), (31) we obtain the following
formula
\begin{equation}
\Theta_2 = \psi_2(s) C_1^2(\sigma^i,t) + \psi_3(s) K^i_jK^j_i + 
\frac{C_2(\sigma^i,t)}{\cosh s},
\end{equation}
where 
\[ \psi_2(s) = - \frac{\sinh s}{2\cosh^2s}, 
\]
\[ 
\psi_3(s) = \frac{1}{2}s \cosh s -  \frac{s}{2\cosh s} 
- \psi_1(s)  \ln(2\cosh s) 
\]
\[
+ \frac{s^2\sinh s}{4\cosh^2s}  - \frac{1}{4 \cosh s} \int^s_0 dx  
\frac{x^2}{\cosh^2x}. 
\]
The integral in $\psi_3(s)$ can easily be evaluated numerically. Due to the
consistency conditions, the functions $C_1, C_2$ in formulas (31), (32) are 
not arbitrary, see below. 

The consistency condition (26) with $k=2$ does not give any restrictions ---
it can be reduced to the identity  $0 = 0$. More interesting is the next 
condition, that is the one with $k=3$. Inserting formula (23) for $f_3$ and
calculating necessary integrals over $s$ we find that it can be written in 
the form of the following inhomogeneous equation for $C_1(\sigma^i,t)$ 
\begin{eqnarray}
& \frac{\gamma_1}{K} ( \partial_t C_1 - g^{kr} \vec{X}_{,r}\dot{\vec{X}}
\partial_k C_1 ) - \frac{1}{\sqrt{g}}\partial_j(\sqrt{g} g^{jk}\partial_kC_1)
- K^i_j K^j_i C_1 & \nonumber \\
& = \frac{\pi^2}{24} K^r_r\left((K^i_i)^2 - 3 K^i_jK^j_i\right). &
\end{eqnarray}
We have also used Allen-Cahn equation (30). Equation (33) determines $C_1$ 
provided that we fix initial data for it. Similarly, the consistency 
condition coming from the fourth order ($k=4$) is 
equivalent to the following homogeneous equation for $C_2$
\begin{eqnarray}
& \frac{\gamma_1}{K} ( \partial_t C_2 - g^{kr} \vec{X}_{,r}\dot{\vec{X}}
\partial_k C_2 ) & \nonumber \\
& - \frac{1}{\sqrt{g}} \partial_j(\sqrt{g} g^{jk}\partial_kC_2)
- K^i_j K^j_i C_2  = 0. &
\end{eqnarray}

The formulas (16), (18), (31) and (32) give a whole family of domain 
walls. To obtain one concrete domain wall solution we have to choose initial
position of the auxiliary surface $S$. Its positions at later times are 
determined from Allen-Cahn equation (30). We also have to fix initial 
values of the functions $C_1, C_2$, and to find the corresponding solutions 
of Eqs.(33), (34).  Notice that we are not allowed to choose the 
initial profile of the domain wall because the dependence 
on the transverse coordinate $s$ is explicitly given. It is known from 
formulas (18), (31) and (32).  
Any choice of the initial data gives an approximate domain wall solution. 
Of course such a choice should not lead to large perturbative corrections 
at least in certain finite time interval. Therefore one should require that
at the initial time $\xi_m C_1 \ll 1 , \xi_m^2 C_2 \ll 1, \xi_m K^i_j \ll 1$. 
The domain wall is located close to the surface $S$ because for large 
$|s|$ the perturbative contributions vanish and the leading term 
$2\arctan(e^s)$ is close to one of the vacuum values $0, \pi$.

Let us remark that Eqs.(30), (33) and (34) imply that a planar 
domain wall ($K^i_j = 0$) can not move, in contradistinction with 
relativistic domain walls for which uniform, inertial motions are possible.

In the presented approach we describe evolution of the domain wall in terms of
the surface $S$ and of the functions $C_1, C_2$. These functions can be 
regarded as fields defined on $S$. In some cases Eqs. (30), (33), (34) can
be solved analytically, one can also use numerical methods. Anyway, these
equations are much simpler than the initial Eq.(5).

The presented formalism is invariant with respect to changes of coordinates
$\sigma^1, \sigma^2$ on $S$. In particular, in a vicinity of any point 
$\vec{X}$ of $S$ we can choose the coordinates in such a way that 
$g_{ik} =\delta_{ik}$ at $\vec{X}$. In these coordinates Eq.(30) has the 
form 
\begin{equation}
\frac{\gamma_1}{K} v = \frac{1}{R_1} + \frac{1}{R_2},
\end{equation}
where $v$ is the velocity in the direction $\vec{p}$ perpendicular to $S$
at the point $\vec{X}$, and $R_1, R_2$ are the main curvature radia of $S$ at
that point. 

As an example, let us consider cylindrical and spherical domain walls.
If  $S$ is a straight cylinder of  radius $R$ then $R_1 = \infty, 
\;R_2 = - R(t)$, $v=\dot{R}$ and Eq.(35) gives 
\begin{equation}
R(t) = \sqrt{R^2_0 - \frac{2K}{\gamma_1}(t-t_0)},    
\end{equation}
where $R_0$ is the initial radius. 
The origin of the Cartesian coordinate frame is located on the symmetry axis 
of the cylinder $S$ (which is the z-axis), $\vec{p}$ is the outward normal to 
$S$, and   $s = (\sqrt{r^2-z^2}-R(t))/\xi_m$, where $r$ is the radial 
coordinate in $R^3$. As $\sigma^1, \sigma^2$ we take the usual cylindrical 
coordinates $z, \phi$. Equations (33), (34) reduce to
\begin{equation}
\frac{\gamma_1}{K} \partial_t C_1 - \left( \partial^2_z C_1 +\frac{1}{R^2}
\partial^2_{\phi}C_1\right)
- \frac{1}{R^2} C_1 =  \frac{\pi^2}{12}  \frac{1}{R^3},  
\end{equation}
\begin{equation}
\frac{\gamma_1}{K} \partial_t C_2 - \left( \partial^2_z C_2 +\frac{1}{R^2}
\partial^2_{\phi}C_2\right) - \frac{1}{R^2} C_2 = 0.  
\end{equation}

If $C_1, C_2$ at the initial time $t_0$ have just constant values 
$C_1(0), C_2(0)$ on the cylinder, then 
\begin{equation}
C_1(t) = \frac{\pi^2}{12 R(t)} \ln(R_0/R(t)) + \frac{R_0}{R(t)} C_1(0), \;\;
C_2(t) = \frac{R_0}{R(t)} C_2(0).
\end{equation}
General solutions of Eqs.(37), (38) can be 
found by splitting $C_1, C_2$ into Fourier modes, but we shall not present
them here. 

The case of spherical domain wall is quite similar. Now $S$ is a sphere of 
radius $R$ and $R_1=R_2 = - R$, $ v = \dot{R}$. Equation (35) gives
\begin{equation}
R(t) = \sqrt{R^2_0 - \frac{4K}{\gamma_1}(t-t_0)},    
\end{equation}
Now the origin is located at the center of the sphere, 
$s = (r - R(t))/\xi_m$, 
and $\vec{p}$ is the outward normal to $S$. As $\sigma^k$ we take the usual
spherical coordinates. Then, Eqs.(33), (34) can be written in the form
\begin{equation}
\frac{\gamma_1}{K} \partial_t C_1 - \frac{1}{R^2}\left(\frac{1}{\sin\theta}
\partial_{\theta}(\sin\theta \partial_{\theta}C_1) + \frac{1}{\sin^2\theta}
\partial^2_{\phi}C_1 \right) - \frac{2}{R^2}C_1 = \frac{\pi^2}{6} 
\frac{1}{R^3},
\end{equation}
and 
\begin{equation}
\frac{\gamma_1}{K} \partial_t C_2 - \frac{1}{R^2}\left(\frac{1}{\sin\theta}
\partial_{\theta}(\sin\theta \partial_{\theta}C_2) + \frac{1}{\sin^2\theta}
\partial^2_{\phi}C_2 \right) - \frac{2}{R^2}C_2 = 0.
\end{equation}
General solution of these equations can be obtained by expanding $C_1, C_2$
into spherical harmonics. In the particular case when $C_1, C_2$ are constant
on the sphere $S$ the solutions $C_k(t)$ have the same form (39) as in the 
previous case except that now  $R(t)$ is given by formula (40). 

In the both cases our approximate formulas are expected to be meaningful 
as long as $R(t)/\xi_m \gg 1$. 

Because we know the transverse profile of the domain wall, we can express
the total free energy $F$ by geometric characteristics of the domain wall.
One should insert our 
approximate solution for $\Theta$ in formulas (2) and (3) for ${\cal F}_e$ 
and ${\cal F}_m$, respectively, and to perform integration over $s$. The 
volume element $d^3x$ is taken in the form
\[
d^3x = \xi_m \sqrt{G} d^2\sigma ds.
\]
For simplicity, let us consider curved domain walls for which 
\[
C_1= 0 = C_2
\]
at the time $t_0$. Straightforward calculation gives
\[
F = - \frac{K}{2}\frac{V}{\xi_m^2} + \frac{2K}{\xi_m}  |S| \;\;\;\;\;\;\;\;\;
\]
\begin{equation}
- \frac{\pi^2}{6} K \xi_m \int d^2\sigma \sqrt{g} (\frac{1}{R_1^2}
+\frac{1}{R_2^2} - \frac{1}{R_1R_2}) + \mbox{terms of the order}\;\xi_m^3,
\end{equation}
where $|S|$ denotes the area of the surface $S$, and $V$ is the total volume
of the liquid crystalline sample. The first  term on the 
r.h.s. of this formula is just a bulk term which appears because the smallest
value of the magnetic free energy density has been chosen to be equal to
$-K/(2\xi_m^2)$. The proper domain wall contribution starts from the second 
term. This term gives the main contribution of the domain wall to $F$. One can
think about the coresponding constant free energy  $2K/\xi_m$ per unit area.
The third term on the r.h.s. of formula (43) represents
the first perturbative correction. It is of the order $(\xi_m/R_i)^2$ when
compared with the main term,  and within the region of validity of our 
perturbative scheme it is small. One can easily show that this term is 
negative or zero.  Hence, it slightly diminishes the total free energy. 
In this sense, the domain walls have negative rigidity --- bending them 
without stretching (i.e., with $|S|$ kept constant) diminishes the free 
energy. 
 
\section{Remarks}
We would like to add several remarks about the expansion in width
and the approximate domain wall solutions it yields. 

\noindent 1. In the presented approach dynamics of the curved domain wall
in the three dimensional space is described in terms of the comoving surface
$S$ and of the functions $C_k$, $k \leq 1$, defined on $S$.  The profile of 
the domain wall has been explicitly expressed by these functions, by 
the transverse coordinate $\xi$, and by the geometric characteristics of $S$.
The surface $S$ and the functions $C_k$ obey equations (30), (33), (34) which 
do not contain $\xi$.  In particular cases these equations can be solved 
analytically, and in general one can look for numerical solutions. Such 
numerical analysis is much simpler than it would be in the case of the 
initial equation (5) for the angle $\Theta$, precisely because one 
independent variable has been eliminated.

\noindent 2. We have used $\xi_m$ as a formal expansion parameter. This may
seem unsatisfactory because it is a
dimensionful quantity, hence it is hard to say whether its value is small 
or large. What really matters is smallness of the corrections 
$\xi_m \Theta_1,\;  \xi_m^2 \Theta_2$. This is the case if 
$\xi_m C_1 \ll 1, \; \xi_m^2 C_2 \ll 1$ and $\xi_m K^i_j \ll 1$, as it follows
from formulas (31) and (32).

\noindent 3. Notice that an assumption that $S$ coincides with the core
for all times in general would not be compatible with the expansion in width. 
If we assume that $C_1 = 0 = C_2$ at certain initial time $t_0$, Eq.(33) 
implies that $C_1 \neq 0$ at later times (unless the r.h.s. of it happens to
vanish).
Then, it follows from formulas (16), (18) and (31) that  $\Theta \neq \pi/2$
at $s=0$, that is on $S$.

\noindent 4. In the present work we have neglected effects which could come 
from perturbations of the exponentially small tails of the domain wall. 
For example, consider a domain wall in the form of infinite 
straight cylinder flattened from two opposite sides. Its front and rear flat
sides have zero curvatures, and according to  Eq.(35) they do not move. 
In our approximations the domain wall shrinks from the sides where the mean 
curvature $1/R_1 + 1/R_2$ does not vanish. Now, in reality the front and 
rear parts interact with each other. This interaction is exponentially small 
only if the two flat parts are far from each other. We have neglected it 
altogether assuming the  $2\arctan(e^s)$ asymptotics at large $s$. In this
sense, our approximate solution takes into account only the effects of 
curvature. 

\noindent 5. Finally, let us mention that the dynamics of domain walls in 
nematic liquid crystals can also be investigated with the help of another
approximation scheme called the polynomial approximation. In the first 
paper \cite{11} it has been applied to a cylindrical domain wall, and in the
second one to a planar soliton. Comparing the two approaches,  the 
polynomial approximation is much cruder than the expansion in width. 
It also contains more arbitrariness, e. g., in choosing concrete form of 
boundary conditions at $|s| \rightarrow \infty$. On the other hand, that 
method is much simpler. It can be useful for rough estimates.

\end{document}